\documentclass{article}
\pdfoutput=1

\usepackage{arxiv}

\usepackage{cite}
\usepackage{amsmath,amssymb,amsfonts}
\usepackage{graphicx}
\usepackage{textcomp}
\usepackage{xcolor}

\usepackage{todonotes}
\usepackage{forest}  % nice tree diagrams within tikz
\usepackage{booktabs}  % for toprule, midrule and other table formatting commands
\usepackage{multirow}  % multirow cells in tables
\usepackage[hidelinks, bookmarks=false]{hyperref}  % \autoref, hidelinks removes colour and border
\usepackage{algpseudocode}  % algorithmic environment and customisation of line numbers
\usepackage{capt-of}  % for \captionof
\usepackage{mathtools}  % for \DeclarePairedDelimiter (https://tex.stackexchange.com/a/118217)
\usepackage[defaultlines=2,all]{nowidow}  % prevent widow and orphan lines
\usepackage{url}  % helpful for formatting URLs (especially around line breaks), including within references
\usepackage[caption=false,font=footnotesize]{subfig}  % options as specified by IEEE (note: can't use subcaption with IEEEtran)
\usepackage{pgfplots}
\usepackage{tikz}
\usepackage{siunitx}

\usetikzlibrary{shapes.misc, matrix, decorations.pathreplacing}  % shapes for star, matrix for "matrix of nodes", decorations for large curly brace
\usepgfplotslibrary{statistics}

\newcommand{\newgdname}{GreedyGD}
\newcommand{\gdINFOCOM}{GD-INFO}
\newcommand{\gdINFOCOMplus}{GD-INFO+}
\newcommand{\gdGLEAN}{GD-GLEAN}
\newcommand{\gdGLEANplus}{GD-GLEAN+}

\newcommand{\algoNameBaseTree}{\texttt{BaseTree}}
\newcommand{\algoNameGreedySelect}{\texttt{GreedySelect}}
\newcommand{\basebits}{B}
\newcommand{\cost}{\mathcal{C}}

\newcommand{\maxdev}{\boldsymbol{\Delta}}
\newcommand{\maxdevzero}{\Delta_{i}^{(0)}}
\newcommand{\nblocal}{n_{b,i}}

\DeclarePairedDelimiter\ceil{\lceil}{\rceil}

\newcommand{\dataResultsCompression}{data/results_compression.csv}
\newcommand{\dataResultsClusteringAR}{data/results_clustering_ar.csv}
\newcommand{\dataResultsClusteringADRGLEAN}{data/results_clustering_ar_vs_adr_GD-GLEAN.csv}
\newcommand{\dataResultsClusteringADRGLEANplus}{data/results_clustering_ar_vs_adr_GD-GLEAN+.csv}
\newcommand{\dataResultsClusteringADRGreedy}{data/results_clustering_ar_vs_adr_GD-greedy.csv}
\newcommand{\dataResultsSamplingCR}{data/results_sampling_compression.csv}
\newcommand{\dataResultsRuntimeVsDimensionality}{data/results_runtime_vs_d_uci-gas_turbine_emissions.csv}

\definecolor{gridcolour}{HTML}{e6e6e6}
\definecolor{annotationsgray}{HTML}{7a7a7a}
\definecolor{colourAlgoComment}{HTML}{787878}

\definecolor{colour1}{HTML}{006ed4}  % this is not the same as Matplotlib (more vibrant)
\definecolor{colour2}{HTML}{ff7f0e}
\definecolor{colour3}{HTML}{2ca02c}
\definecolor{colour4}{HTML}{d62728}
\definecolor{colour5}{HTML}{9467bd}
\definecolor{colour6}{HTML}{8c564b}
\definecolor{colour7}{HTML}{e377c2}
\definecolor{colour8}{HTML}{7f7f7f}
\definecolor{colour9}{HTML}{bcbd22}
\definecolor{colour10}{HTML}{17becf}

\newcommand{\algowidth}{0.94\columnwidth}
\algnewcommand{\LineComment}[1]{\textcolor{colourAlgoComment}{// \textit{#1}}}  % custom full line comment
\newcounter{algoCounter}  % custoom counter for algorithms
\newcommand{\showAlgoCounter}[2]{\noalign{\refstepcounter{algoCounter}#1}\textbf{Algorithm~\thealgoCounter} #2}  % command to increment and print algorithm counter, the \noalign is necessary to remove vertical space automatically added by \hyperref

\tikzset{
	cross/.style={
		cross out,
		draw,
		minimum size=2*(#1-\pgflinewidth),
		inner sep=0pt,
		outer sep=0pt},
}
\pgfdeclareplotmark{mystar}{
	\draw[white,fill=white] circle (3.3pt);
	\draw node[cross=3.0pt,very thick] {};
}

\tikzset{
	every pin/.style={
		pin edge={black,thick},
		font=\footnotesize
	},
	bits table style/.style={
		matrix of nodes,
		row sep=-\pgflinewidth,
		column sep=-\pgflinewidth,
		nodes={draw=black, font=\ttfamily},
	}
}

\newcommand{\graphStandardWidth}{0.75\columnwidth}
\newcommand{\graphStandardHeight}{0.65\columnwidth}

\pgfplotsset{
	compat=1.17,
	every axis/.style={
		font=\footnotesize,
		table/col sep=comma,
		xtick pos=bottom,
		ytick pos=left,
		tick align=outside,
		title style={yshift=-0.8em},
		legend cell align={left},
		width=\graphStandardWidth, height=\graphStandardHeight,
	},
	/pgf/declare function={Floor(\x) = round(\x-0.49);},  % accurate floor function
	plain line plot style/.style={
		cycle list={
			{black!90, mark options={draw=black,scale=1.0}, mark=x},
			{black!90, mark options={draw=black,scale=1.0}, mark=o},
			{black!90, mark options={draw=black,scale=1.0}, mark=star},
			{black!90, mark options={draw=black,scale=1.0}, mark=square},
			{black!90, mark options={draw=black,scale=1.0}, mark=triangle},
			{black!90, mark options={draw=black,scale=1.0}, mark=diamond},
			{black!90, mark options={draw=black,scale=1.0}, mark=otimes},
			{black!90, mark options={draw=black,scale=1.0}, mark=+},
		},
		line width=0.5pt,
		legend style={
			align=left,
			column sep=0.1em,
			/tikz/every odd column/.style={yshift=0.1em},
			/tikz/nodes={inner sep=0.1em},
		},
	},
	custom line plot style/.style={
		cycle list={
			{colour1!40, mark options={draw=colour1!70, scale=1.4}, mark=*},
			{colour2!40, mark options={draw=colour2!70, scale=1.25}, mark=square*},
			{colour3!40, mark options={draw=colour3!70, scale=1.8}, mark=triangle*},
			{colour4!40, mark options={draw=colour4!70, scale=1.4}, mark=pentagon*},
			{colour5!40, mark options={draw=colour5!70, scale=1.6}, mark=diamond*},
			{colour6!40, mark options={draw=colour6!70, scale=1.4}, mark=oplus*},
			{colour7!40, mark options={draw=colour7!70, scale=1.4}, mark=10-pointed star},
			{colour8!40, mark options={draw=colour8!70, scale=1.25}, mark=halfsquare*}
		},
		line width=0.5pt,
		legend style={
			align=left,
			column sep=0.1em,
			/tikz/every odd column/.style={yshift=0.1em},
			/tikz/nodes={inner sep=0.1em},
		},
	},
	custom scatter plot style/.style={
		only marks,
		cycle list={
			{colour1, mark options={scale=1.4, line width=0.5pt, fill opacity=0.45, draw opacity=0.7}, mark=*},
			{colour2, mark options={scale=1.25, line width=0.5pt, fill opacity=0.45, draw opacity=0.7}, mark=square*},
			{colour3, mark options={scale=1.8, line width=0.5pt, fill opacity=0.45, draw opacity=0.7}, mark=triangle*},
			{colour4, mark options={scale=1.4, line width=0.5pt, fill opacity=0.45, draw opacity=0.7}, mark=pentagon*},
			{colour5, mark options={scale=1.6, line width=0.5pt, fill opacity=0.45, draw opacity=0.7}, mark=diamond*},
			{colour6, mark options={scale=1.4, line width=0.5pt, fill opacity=0.45, draw opacity=0.7}, mark=oplus*},
			{colour7, mark options={scale=1.4, line width=0.5pt, fill opacity=0.45, draw opacity=0.7}, mark=10-pointed star},
			{colour8, mark options={scale=1.25, line width=0.5pt, fill opacity=0.45, draw opacity=0.7}, mark=halfsquare*}
		},
	},
	dual boxplot style/.style={
		boxplot={
			draw direction=y,
			draw position={1/3 + Floor(\plotnumofactualtype/2) + 1/3*mod(\plotnumofactualtype,2)},
			box extend=0.3,
			every median/.style={very thick},
		},
		cycle list={
			{black, mark=*, solid, fill=colour1, fill opacity=0.25},
			{black, mark=*, solid, fill=colour2, fill opacity=0.25}
		},
		x tick label as interval,
		legend style={
			align=left,
			column sep=0.2em,
			/tikz/every odd column/.style={yshift=0.1em},
			/tikz/nodes={inner sep=0.1em},
		},
	},
	single boxplot style/.style={
		boxplot={
			draw direction=x,
			box extend=0.7,  % decrease width of boxes
			every median/.style={very thick},
			mark=o,
		},
		every boxplot/.style={color=black, fill=colour1, solid, fill opacity=0.25},
	}
}

\hypersetup{
    pdftitle={\newgdname: Enhanced Generalized Deduplication for Direct Analytics in IoT},
    pdfauthor={Aaron Hurst, Daniel E. Lucani, and Qi Zhang},
    pdfkeywords={Data Compression, Data Analytics, Generalized Deduplication, Clustering, IoT},
}

\begin{document}

    \title{\newgdname{}: Enhanced Generalized Deduplication for Direct Analytics in IoT}
    \date{April 13, 2023}
    \author{
        Aaron Hurst, Daniel E. Lucani and Qi Zhang\\
        DIGIT Centre and the Department of Electrical and Computer Engineering \\
        Aarhus University, Aarhus, Denmark \\
        E-mail: {\texttt{\{ah,daniel.lucani,qz\}@ece.au.dk}}
    }
    \renewcommand{\shorttitle}{GreedyGD}
    \maketitle

	\begin{abstract}
	Exponential growth in the amount of data generated by the Internet of Things currently pose significant challenges for data communication, storage and analytics and leads to high costs for organisations hoping to leverage their data.
	Novel techniques are therefore needed to holistically improve the efficiency of data storage and analytics in IoT systems.
	The emerging compression technique Generalized Deduplication (GD) has been shown to deliver high compression and enable direct compressed data analytics with low storage and memory requirements.
	In this paper, we propose a new GD-based data compression algorithm called \newgdname{} that is designed for analytics.
	Compared to existing versions of GD, \newgdname{} enables more reliable analytics with less data, while running 11.2$ \times $ faster and delivering even better compression.
\end{abstract}

% IEEE keywords sourced from here: https://www.ieee.org/content/dam/ieee-org/ieee/web/org/pubs/taxonomy_v101.pdf
\keywords{Data Compression, Data Analytics, Generalized Deduplication, Clustering, IoT}

	\section{Introduction}

Sensor data collected via the Internet of Things (IoT) are increasingly relied upon to make critical business decisions, improve efficiency and keep people safe~\cite{Kumar_2019}.
However, managing these data in a traditional, centralised manner becomes costly not only in terms of storage and compute, but also latency, as data generation outpaces improvements in hardware. % , especially for organisations utilising cloud-based services.
Likewise, Edge servers, while able to reduce communication costs and latency~\cite{Yu_2018}, are also limited by their lower computing power and higher storage unit-cost. % also Nayak_2021, Liu_2019a
Overall, the cost of transmitting, storing and analysing IoT data may limit the utility and sustainability of IoT systems by forcing organisations to discard data or retain it only for a short time~\cite{Sisinni_2018,Lee_2019}.
Thus, there is a need for consolidated frameworks that reduce the cost of IoT data storage, while improving analytics latency.

\begin{figure}[!t]
	\centering
	\footnotesize
\begin{tikzpicture}
    % Matrix of data chunks
    \matrix [
    bits table style,
    column 1/.style={nodes={fill=colour1!15}},
    column 2/.style={nodes={fill=colour1!15}},
    column 3/.style={nodes={fill=colour1!15}},
    column 8/.style={nodes={fill=colour1!15}},
    label={[yshift=-0.1em]above:Data chunks},
    ] (chunks)
    {
        1 & 0 & 1 & 0 & 0 & 0 & 0 & 0 \\
        1 & 1 & 1 & 0 & 0 & 0 & 1 & 0 \\
        1 & 0 & 1 & 1 & 0 & 1 & 1 & 0 \\
        1 & 1 & 1 & 1 & 1 & 0 & 0 & 0 \\
        1 & 1 & 1 & 0 & 0 & 0 & 1 & 0 \\
        1 & 1 & 0 & 0 & 0 & 0 & 0 & 0 \\
        1 & 1 & 1 & 1 & 1 & 1 & 1 & 0 \\
    };

    % Matrix of deviations
    \matrix [
    bits table style,
    label={[yshift=-0.1em]above:Deviations},
    anchor=west,
    xshift=2.6em,
    ] (deviations) at (chunks.east)
    {
        0 & 0 & 0 & 0 \\
        0 & 0 & 0 & 1 \\
        1 & 0 & 1 & 1 \\
        1 & 1 & 0 & 0 \\
        0 & 0 & 0 & 1 \\
        0 & 0 & 0 & 0 \\
        1 & 1 & 1 & 1 \\
    };

    % Base IDs
    \matrix [
    bits table style,
    nodes={fill=colour2!15},
    label={[yshift=-0.1em]above:IDs},
    anchor=north west,
    xshift=0.02em,
    ] (baseIDs) at (deviations.north east)
    {
        0 \\
        1 \\
        0 \\
        1 \\
        1 \\
        2 \\
        1 \\
    };

    % Matrix of bases
    \matrix [
    bits table style,
    nodes={fill=colour1!15},
    label={[yshift=-0.1em]above:Bases},
    anchor=north west,
    xshift=0.3em,
    ] (bases) at (baseIDs.north east)
    {
        1 & 0 & 1 & 0 \\
        1 & 1 & 1 & 0 \\
        1 & 1 & 0 & 0 \\
    };

    % Base counts
    \matrix [
    bits table style,
    nodes={fill=colour2!15},
    label={[yshift=-0.1em]above:Counts},
    anchor=north west,
    xshift=0.02em,
    ] (baseCounts) at (bases.north east)
    {
        2 \\
        4 \\
        1 \\
    };

    % Bounding boxes
    \node[
    rectangle,
    draw=annotationsgray,
    dashed,
    minimum width=11.6em,
    minimum height=12.3em,
    yshift=0.8em,
    ] (uncompressed) at (chunks) {};
    \node[
    rectangle,
    draw=annotationsgray,
    draw,
    dashed,
    minimum width=17.3em,
    minimum height=12.3em,
    xshift=1.3em,
    anchor=west,
    ] (compressed) at (uncompressed.east) {};

    % Bounding box labels
    \node[annotationsgray, anchor=center, yshift=-0.1em, fill=white] () at (uncompressed.north) {Uncompressed data};
    \node[annotationsgray, anchor=center, yshift=-0.1em, fill=white] () at (compressed.north) {Compressed data};
\end{tikzpicture}
%	\vspace{0.0cm}
	\caption{
		GD applied to 8-bit data chunks.
		The first three bits and last bit (highlighted in blue) are allocated to the base and deduplicated, while the remaining bits are the deviations and are stored verbatim alongside an ID linking them to the relevant bases.
	}
	\label{fig:gd_illustration}
	\vspace{-0.5em}
\end{figure}
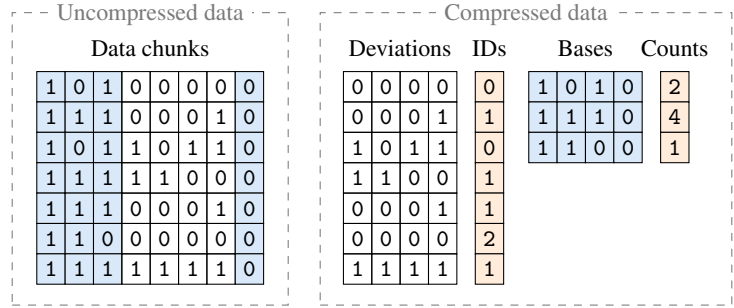

The emerging compression technique Generalized Deduplication (GD)~\cite{Vestergaard_2019a, Vestergaard_2019, Vestergaard_2020} provides an effective solution to this need.
Namely, it offers high, lossless end-to-end compression in IoT systems, as well as efficient approximate analytics without the need for decompression~\cite{Vestergaard_2020,Hurst_2021,Hurst_2022}.
Moreover, the structure of GD-compressed data allows data to be split between Edge and Cloud servers to optimise for storage cost and analytics latency~\cite{Hurst_2022}.
GD performs compression by splitting data chunks into \textit{bases}, which are deduplicated, and \textit{deviations}, which are stored verbatim (see \autoref{fig:gd_illustration}).
Deviations are stored together with an ID that links them to their associated base, allowing for lossless decompression.
Compressing a dataset with GD involves two stages: 1)~configuration, i.e., deciding which bits to allocate to the base, and 2)~compression, i.e., splitting data chunks into bases and deviations.

%GD benefits:
% - high compression,
% - end-to-end lossless compression,
% - highly efficient approximate analytics without the need for decompression,
% - ability to split compressed data between Cloud (for storage) and the Edge (for analytics)

In this paper, we propose \newgdname{}, which is a GD-based data compression algorithm designed for running analytics directly on compressed data.
Compared to previous versions of GD, it addresses a number of critical pitfalls for compression speed and analytics quality, including configuration time and base-sample order preservation.
\newgdname{} even delivers better compression than existing versions of GD.
The main contributions of this paper include:
\begin{enumerate}
	\item The \newgdname{} compression algorithm, which is designed for direct compressed data analytics,
	
	\item Data preprocessing for GD that improves compression of floating point data,
	
	\item Comprehensive evaluation \newgdname{}'s performance in comparison to existing GD versions and universal lossless compressors in terms of compression ratio, runtime and analytics, and
	
	\item Evaluation of configuring GD using a data subset.
\end{enumerate}

The paper is structured as follows.
\autoref{sec:related_work} discusses related work.
\autoref{sec:nomenclature} details our nomenclature.
\autoref{sec:greedy_gd} describes \newgdname{}.
\autoref{sec:results} evaluates the performance of \newgdname{}. % in terms of preprocessing, compression, runtime, sampling and analytics.
Finally, \autoref{sec:conclusion} concludes the paper.

	\section{Related Work}
\label{sec:related_work}

Data compression algorithms can be either lossy or lossless.
Lossy algorithms typically offer more compression, but at the cost of degraded data fidelity, which may be unsuitable for critical IoT applications.
We therefore focus on lossless compression.
A wide variety of universal lossless compression algorithms are available~\cite{Deutsch_1996a, Facebook_, Google_2021, Seward_2019, Collet_2020}, as well as powerful IoT-specific lossless time series compression algorithms, such as Sprintz~\cite{Blalock_2018}.
%typically based on Huffman~\cite{Huffman_1952} or Lempel-Ziv (LZ) codes~\cite{Ziv_1977} and transforms such as run-length encoding, delta encoding and Burrows-Wheeler~\cite{Burrows_1994}.
%Examples include bzip2~\cite{Seward_2019}, zstd~\cite{Facebook_}, DEFLATE~\cite{Deutsch_1996} and Snappy~\cite{Google_2021}.
%These algorithms offer good compression and often provide the ability to trade-off between compression and speed.
However, existing algorithms are limited in terms of random access (important for analytics), small-chunk performance (relevant for IoT sensors) and their ability to run on resource-limited IoT devices~\cite{Vestergaard_2021}.

GD-based data compression, on the other hand, performs well in these areas and also provides high compression.
To achieve this, the choice of base bits, i.e., the set of bit indices mapped to the bases, is critical and should depend on both data and application. % (e.g. online operation, analytics).
The version of GD described in~\cite{Vestergaard_2020}, which we refer to as \gdINFOCOM{}, selects base bits according to inter-bit correlation.
It first maps all bits to the base and computes the compressed data size.
Next, the bit with lowest inter-bit correlation is moved from the base to the deviation and compressed data size is re-computed.
This continues until the first local minimum for compressed data size.
The base bits at this point are then used for compression.
To limit runtime, configuration is run on at most the first $ \mathrm{10^6} $ samples.

While \gdINFOCOM{} delivers good performance, it has some significant limitations, including: 1) no efficient method for computing the compressed data size, 2) inter-bit correlation is not always an accurate indicator of deduplication potential, 3) terminating at the first local minimum is often sub-optimal, 4) no standard approach for multidimensional data, and 5) poor compressed data analytics performance, as shown in~\cite{Hurst_2022}.
The effect of using a data subset (i.e., the first $ \mathrm{10^6} $ samples) for configuration has also not been investigated.

In~\cite{Hurst_2022}, an analytics-tailored version of GD was proposed, which we refer to as \gdGLEAN{}.
Different to \gdINFOCOM{}, this compressor concedes some compression for improved analytics performance, but otherwise suffers from the same limitations as \gdINFOCOM{}.

Two related algorithms that bear notable similarities to GD are ALACRITY~\cite{Jenkins_2013} and DigitHist~\cite{Shekelyan_2017}.
While both algorithms perform compression by extracting base-like ``bins'' from data, they are limited with respect to GD in that they place significant restrictions on which bits can be assigned to the these bins.
Moreover, in contrast to GD, DigitHist uses lossy compression.

In this paper, we address the aforementioned limitations by proposing \newgdname{}, which delivers significantly more accurate analytics on compressed data, while running faster and providing even better compression.

	\section{Nomenclature}
\label{sec:nomenclature}

% Nomenclature table
\begin{table}[!t]
	\centering
	\renewcommand{\arraystretch}{1.2}
	\caption{Nomenclature}
	\setlength\tabcolsep{0.8em}
    {\small\begin{tabular}{clcl}
	\toprule
	\textbf{Notation} & \textbf{Meaning}    & \textbf{Notation} & \textbf{Meaning}             \\ \midrule
	      $ n $       & No. of data samples &       $ S $       & Compressed data size         \\
	      $ d $       & No. of dimensions   &    $ \Delta $     & Maximum deviation            \\
	     $ n_b $      & No. of bases        &    $ \lambda $    & Balancing factor             \\
	  $ \basebits $   & Set of base bits    &    $ \alpha $     & Exploration factor           \\
	     $ l_c $      & Bits per chunk      &     $ \cost $     & \algoNameGreedySelect{} cost \\
	     $ l_b $      & Bits per base       &        CR         & Compression ratio            \\
	     $ l_d $      & Bits per deviation  &        ADR        & Analytics data ratio         \\
	   $ l_{id} $     & Bits per base ID    &        AR         & Approximation ratio          \\
	   $ l_{bc} $     & Bits per base count &        AMI        & Adjusted mutual info.        \\ \bottomrule
\end{tabular}

}
	\label{tab:nomenclature}
	\vspace{-0.3em}
\end{table}

See \autoref{tab:nomenclature} for a summary of our nomenclature.
We consider a dataset to contain $ n $ samples and $ d $ dimensions.
The set of bit indices that are allocated to bases by GD is referred to as the \textit{base bits} and denoted $ \basebits $.
The total number of (deduplicated) bases is denoted $ n_b $.
A binary data \textit{chunk} is formed by concatenating a data sample across its dimensions and has length~$ l_c $ bits.
The number of bits per base and deviation are denoted $ l_b $ and $ l_d $, respectively.
Hence, $ l_c = l_b + l_d $.
GD stores base IDs and base counts using $ l_{id} = \ceil{\log_2(n_b)} $ and $ l_{bc} = \ceil{\log_2(n)} $ bits each, respectively.

The \textit{maximum deviation}, $ \maxdev $, is the maximum difference between samples mapped to the same base.
For example, the minimum value mapped to first base (all zero deviation bits) in \autoref{fig:gd_illustration} is $ \texttt{10100000} $, or 160, while the maximum value (all one deviation bits) is $ \texttt{10111110} $, or 190.
This gives a maximum deviation of $ \texttt{00011110} $, or 30, which is simply all deviation bits set to~$ \texttt{1} $.
Note, $ \maxdev $ is a vector for multidimensional data and varies between bases for floating point data since the values of mantissa bits depend on the exponent.

The total compressed data size for GD, $ S $, is equal to:
\begin{align} \label{eq:S}
	S = n_b \left(l_b + l_{bc}\right) + n \left(l_{id} + l_d\right) + S_{params},
\end{align}
where $ S_{params} $ is the size of GD's parameters (typically negligible).
Compression performance is measured using the \textit{compression ratio} (CR), defined as follows:
\begin{align} \label{eq:cr}
	\mathrm{CR} = \frac{\text{size of compressed data}}{\text{size of uncompressed data}}.
\end{align}
As such, lower values (below 1) are better.
Additionally, the amount of (compressed) data that must be accessed in order to run analytics directly on GD-compressed data is measured using the \textit{analytics data ratio} (ADR), which is equal to:
\begin{align} \label{eq:adr}
	\mathrm{ADR} = \frac{\text{size of data used for analytics}}{\text{size of uncompressed data}} \approx \frac{n_b (l_b + l_{bc})}{nl_c}.
\end{align}

We use $ k $-means clustering~\cite{Hartigan_1979} as a test case for the analytics performance of \newgdname{}.
Three performance metrics are used.
The first, approximation ratio (AR), is defined in terms of the sum of squared errors (SSE) as follows:
\begin{align}
	\mathrm{AR} = \frac{\text{clustering SSE using compressed data}}{\text{clustering SSE using uncompressed data}},
\end{align}
As such, $ \mathrm{AR} \ge 1 $ and lower values indicate better performance.
A value of 1 indicates that no loss in clustering quality due to running clustering on compressed data instead of uncompressed data.

The second metric we use is adjusted mutual information (AMI), which is an information theoretic measure of the similarity between two clusterings.
%This metric is based on the ratio between the mutual information between two clusterings and the product of their entropies and adjusted to remove bias towards higher numbers of clusters~\cite{Vinh_2009}.
In this case, AMI is always between 0 and 1 and higher values are better, i.e., indicate more similar clusterings.

The third metric is the Silhouette coefficient, which measures how well data points fit with those in the same cluster compared to those in the next most appropriate cluster. %~\cite{Rousseeuw_1987}.
More formally, for a given data point $ \boldsymbol{x} $, if $ a(\boldsymbol{x}) $ is the mean distance between $ \boldsymbol{x} $ and other points in the same cluster and $ b(\boldsymbol{x}) $ is the mean distance between $ \boldsymbol{x} $ and points in the next best cluster, then the silhouette of $ \boldsymbol{x} $ is:
\begin{equation}
	s(\boldsymbol{x}) = \frac{b(\boldsymbol{x}) - a(\boldsymbol{x})}{\max \left( a(\boldsymbol{x}), b(\boldsymbol{x})  \right)}.  \label{eq:silhouette}
\end{equation}
The Silhouette coefficient for an entire clustering is the mean of the silhouettes of all data points.
The silhouette coefficient ranges between $ \pm1 $ such that higher values are better and values close to 1 indicate dense, well-separated clusters.

%Before continuing, we briefly note that we handle multidimensional data by concatenating data samples across dimensions such that, for $ d $ dimensions and 32 bit values, data chunks will consist of $ 32d $ bits.
%This approach allows effective cross-column analytics such as clustering~\cite{Hurst_2021}.
%In future work, we intend to investigate column-wise compression.

	\section{\newgdname{} Algorithm}
\label{sec:greedy_gd}

\begin{figure}[!t]
	\centering
	\footnotesize
	\renewcommand{\arraystretch}{1.1}
	\begin{tabular}{p{\algowidth}}
	\toprule
	\showAlgoCounter{\label{algo:greedygd}}{\newgdname{}} \\ \midrule
	\textbf{Inputs:} dataset $ \mathcal{D} $, exploration factor $ \alpha $, balancing factor $ \lambda $ \\
	\textbf{Outputs:} compressed dataset $ \mathcal{D}_{GD} $ \\
	\vspace{-0.65em}
	\begin{algorithmic}[1]
		\State $ \mathcal{D}' \leftarrow $ preprocess $ \mathcal{D} $ \Comment{\autoref{sec:greedy_gd:preprocessing}}
		\State $ D \leftarrow $ extract subset from $ \mathcal{D}' $ \Comment{\autoref{sec:greedy_gd:sampling} (optional)}
		\State $ \basebits \leftarrow $ \algoNameGreedySelect{}$ ( D', \alpha, \lambda ) $  \Comment{\autoref{sec:greedy_gd:greedy_select}}
		\State Apply GD to $ \mathcal{D}' $ according to $ \basebits $ 
%		\State \Return $ \mathcal{D}_{GD} $
	\end{algorithmic}
	\\[-1em] \bottomrule
\end{tabular}

\end{figure}

The key to GD-based compression is selecting an effective set of base bits.
This is challenging due to the number of possible combinations.
%For the simple example in \autoref{fig:gd_illustration}, there are $ 2^8 $ or 256 possible configurations.
For example, d-dimensional 32-bit data has $ 2^{32d} $ possible configurations, i.e., over 4 billion \textit{per dimension}.
Enumerating all of these is infeasible since one must count the number of unique bases for each possible combination, which is highly time consuming.

\gdINFOCOM{} solves this problem by ordering bits according to inter-bit correlation and terminating at the first local minima for compressed data size.
This limits the number of configurations enumerated to at most $ l_c $.
However, while this makes \gdINFOCOM{} computationally feasible, the aforementioned limitations make it suboptimal.

The proposed \newgdname{} compression algorithm, outlined at a high level in Algorithm~\ref{algo:greedygd}, aims to find better base bit sets and evaluate them more quickly.
In a nutshell, \newgdname{} works by 1)~applying novel data preprocessing, 2)~optionally selecting a subset of the data to accelerate configuration, 3)~selecting the base bits using a greedy algorithm and 4)~compressing the data using GD.

The remainder of this section is structured as follows.
\autoref{sec:greedy_gd:base_tree} describes the \algoNameBaseTree{} algorithm, which is used to quickly count bases.
\autoref{sec:greedy_gd:greedy_select} describes the \algoNameGreedySelect{}, which builds on \algoNameBaseTree{} to select high quality base bits.
\autoref{sec:greedy_gd:preprocessing} details how preprocessing can improve compression, while \autoref{sec:greedy_gd:sampling} details how using data subsets can improve runtime.
Finally, \autoref{sec:greedy_gd:complexity} analyses the time complexity of \newgdname{}.

\subsection{Counting bases}
\label{sec:greedy_gd:base_tree}

% BaseTree illustration
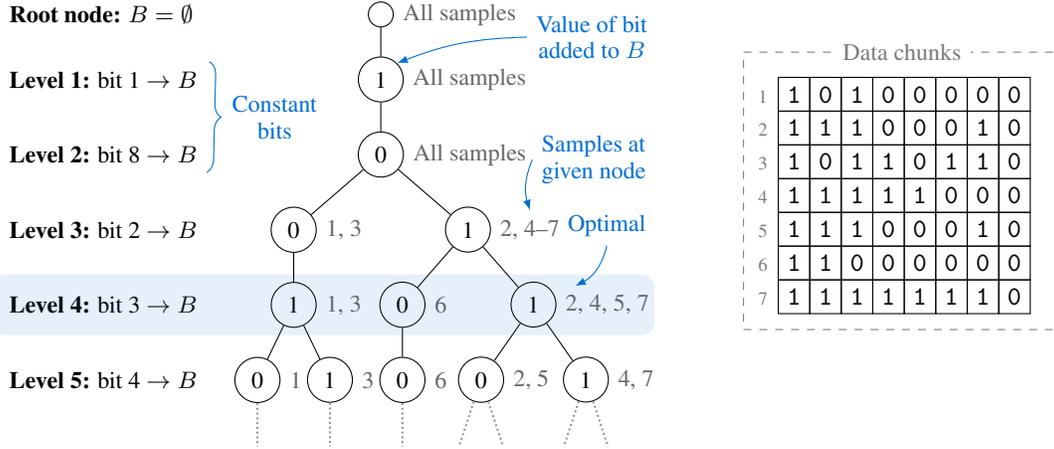
\begin{figure}[!t]
	\centering
	\tikzset{
	levels style/.style={
		align=left,
		anchor=west
	}
}

\tikzset{
	braces style/.style={
		decorate,
		decoration={brace,amplitude=0.5em},
		colour1,
	}
}

\begin{forest}
	/tikz/every node/.append style={font=\footnotesize},
	for tree = {
		circle,
		draw,
		s sep=1.0em,
	},
	where level=6{
		edge={black!40,densely dotted,thick},
		draw=none,
	}{},
	[,name=root
	[1,name=level1
	[0,name=level2
		[0,name=level3
			[1,name=level4
				[0,name=level5
					[]
				]
				[1,name=level5n1
					[]
				]
			]
		]
		[1,name=level3n1
			[0,name=level4n1
				[0,name=level5n2
					[]
				]
			]
			[1,name=level4n2
				[0,name=level5n3
					[][]
				]
				[1,name=level5n4
					[][]
				]
			]
		]
	]
	]
	]
	%
	% Tree level annotations
	\node[left=2.2cm of root, align=left, anchor=east] (rootlabel) {\textbf{Root node:} $ \basebits = \emptyset $};
	\node[levels style] (L1) at (level1 -| rootlabel.west) {\textbf{Level 1:} bit 1 $ \rightarrow \basebits $};
	\node[levels style] (L2) at (level2 -| rootlabel.west) {\textbf{Level 2:} bit 8 $ \rightarrow \basebits $};
	\node[levels style] (L3) at (level3 -| rootlabel.west) {\textbf{Level 3:} bit 2 $ \rightarrow \basebits $};
	\node[levels style] (L4) at (level4 -| rootlabel.west) {\textbf{Level 4:} bit 3 $ \rightarrow \basebits $};
	\node[levels style] (L5) at (level5 -| rootlabel.west) {\textbf{Level 5:} bit 4 $ \rightarrow \basebits $};
	\draw[braces style] (L1.north east)+(0.1em,0) -- (L2.south east)+(-0.1em,0) node [midway, anchor=west, align=center, xshift=2mm] {Constant\\bits};
	%
	% Annotations of where samples end up
	\node[black!60, anchor=west] () at (root.east) {All samples};
	\node[black!60, anchor=west] () at (level1.east) {All samples};
	\node[black!60, anchor=west] () at (level2.east) {All samples};
	\node[black!60, anchor=west] () at (level3.east) {1, 3};
	\node[black!60, anchor=west] (3n1samples) at (level3n1.east) {2, 4--7};
	\node[black!60, anchor=west] () at (level4.east) {1, 3};
	\node[black!60, anchor=west] () at (level4n1.east) {6};
	\node[black!60, anchor=west] () at (level4n2.east) {2, 4, 5, 7};
	\node[black!60, anchor=west] () at (level5.east) {1};
	\node[black!60, anchor=west] () at (level5n1.east) {3};
	\node[black!60, anchor=west] () at (level5n2.east) {6};
	\node[black!60, anchor=west] () at (level5n3.east) {2, 5};
	\node[black!60, anchor=west] () at (level5n4.east) {4, 7};
	%
	% Annotation for meaning of node labels (i.e. value of bit)
	\node[colour1, align=center, xshift=2.5cm, yshift=1.6em] (meaning) at (level1.east) {Value of bit\\added to $ \basebits $};
	\draw [-latex, colour1, solid] (meaning.west)+(0em,0em) to [bend right=10] (level1.north east); %+(-5.4em,-2.15em);
	%
	% Annotate around optimal configuration
	\node[colour1, rounded corners=0.5em, fill=colour1, fill opacity=0.1, minimum width=8.8cm, minimum height=2.3em, anchor=west, xshift=-0.3em] (optimal) at (L4.west) {};
	\node[colour1, anchor=east, xshift=0cm, yshift=3em] (optimalnote) at (optimal.east) {Optimal};
	\draw [-latex, colour1, solid] (optimalnote.south) to [bend left=25] +(-1.2em,-1.53em);
	%
	% Annotation for samples that reach a given node
	\node[colour1, anchor=east, align=center, xshift=0cm, yshift=2.5em] (samples) at (optimalnote.east) {Samples at\\given node};
	\draw [-latex, colour1, solid] (samples.west) to [bend right=20] (3n1samples.north);
	%
	% Matrix of data chunks
	\matrix [
		bits table style,
		anchor=north west,
		xshift=13.5em,
        yshift=-2.5em,
		ampersand replacement=\&,
		column 1/.style={nodes={annotationsgray, draw=none, font=\rmfamily, font=\scriptsize, minimum width=1.2em}},
	] (chunks1) at (root.north)
	{
		1 \& 1 \& 0 \& 1 \& 0 \& 0 \& 0 \& 0 \& 0 \\
		2 \& 1 \& 1 \& 1 \& 0 \& 0 \& 0 \& 1 \& 0 \\
		3 \& 1 \& 0 \& 1 \& 1 \& 0 \& 1 \& 1 \& 0 \\
		4 \& 1 \& 1 \& 1 \& 1 \& 1 \& 0 \& 0 \& 0 \\
        5 \& 1 \& 1 \& 1 \& 0 \& 0 \& 0 \& 1 \& 0 \\
        6 \& 1 \& 1 \& 0 \& 0 \& 0 \& 0 \& 0 \& 0 \\
        7 \& 1 \& 1 \& 1 \& 1 \& 1 \& 1 \& 1 \& 0 \\
	};
	%
	% Bounding box for matrix of nodes
	\node[
		rectangle,
		draw=annotationsgray,
		dashed,
		anchor=north west,
		minimum width=12em,
		minimum height=10.5em,
		yshift=0.6em,
		xshift=0.2em,
	] (chunksbb) at (chunks1.north west) {};
	\node[annotationsgray, anchor=center, fill=white] () at (chunksbb.north) {Data chunks};
\end{forest}
	\caption{
		\algoNameBaseTree{} partially applied to the data chunks from \autoref{fig:gd_illustration}, which are repeated at the bottom of the figure for convenience.
		The highlighted optimal configuration, level 4, corresponds to the allocate of base and deviation bits in \autoref{fig:gd_illustration}.
	}
	\label{fig:basetree_illustration}
\end{figure}

To accelerate counting bases for GD configuration, we propose the \algoNameBaseTree{} algorithm, which is illustrated in \autoref{fig:basetree_illustration}.
The idea is to store the distribution of bases for a given base bits set in a tree structure.
As shown in \autoref{fig:basetree_illustration}, the root node corresponds to $ \basebits = \emptyset $ and contains all data samples.
Each subsequent layer corresponds to an additional bit added to $ \basebits $ and potentially additional nodes, depending on the values of the selected bit.
For example, in level 2, only one node is created because bit 8 is always $ \texttt{0} $ (i.e. constant).
In contrast, in level 3, the data are split between two child nodes because bit 3 contains both $ \texttt{0} $s and $ \texttt{1} $s.
In level 4, only one child is spawned from the left parent node because samples 1 and 3 both have a $ \texttt{1} $ in bit 3, while the right parent node spawns two child nodes because bit 3 contains both $ \texttt{0} $s (left child node) and $ \texttt{1} $s (right child node) among samples 2 and 4--7.

This structure can be used to quickly count the number of bases for a given set of base bits $ \basebits $ since tree width corresponds to $ n_b $.
Counting bases, therefore, simply requires 1)~adding the desired base bits to the tree and 2)~counting the number of leaf nodes.
Note that tree height corresponds to the number of base bits $ l_b $ and root-to-leaf paths correspond to individual bases.

\subsection{Selecting base bits}
\label{sec:greedy_gd:greedy_select}

% Greedy select algorithm
\begin{figure}[!t]
	\centering
	\footnotesize
	\renewcommand{\arraystretch}{1.1	}
	\begin{tabular}{p{\algowidth}}
	\toprule
	\showAlgoCounter{\label{algo:greedyselect}}{\algoNameGreedySelect{}} \\ \midrule
	\textbf{Inputs:} dataset $ \mathcal{D} $, exploration factor $ \alpha $, balancing factor $ \lambda $ \\
	\textbf{Outputs:} base bits $ \basebits $ \\
	\vspace{-0.65em}
	\begin{algorithmic}[1]
		\State \LineComment{Initialisation}
		\State $ \basebits \leftarrow $ constant bits in $ \mathcal{D} $ \label{algo:greedyselect:line:constant_bits_start}
		\State Expand \algoNameBaseTree{} with $ \basebits $ \label{algo:greedyselect:line:constant_bits_end}
		\State $ \basebits_{\mathrm{best}} \leftarrow \basebits $
		\State $ \cost_{\mathrm{best}} \leftarrow \infty$
		
		\State \LineComment{Iteratively add base bits until termination}
		\While{not all bits in $ \basebits $}
		
			\State $ b_{\mathrm{loc}} \leftarrow -1 $  \Comment{reset lowest cost bit for local optimisation}
			\State $ \cost_{\mathrm{loc}} \leftarrow \infty $ \Comment{reset lowest cost for local optimisation}
			
			\For{$ i = 1 $ to number of columns} \label{algo:greedyselect:line:iterate_m_start}
				\State $ b_i \leftarrow $ most significant bit from column $ i $ not in $ \basebits $ \label{algo:greedyselect:line:nextmostsigbit}
				
				\State $ \nblocal \leftarrow n_b $ after adding $ b_i $ to $ \basebits $ \label{algo:greedyselect:line:n_b} \Comment{use \algoNameBaseTree{}}
				
				\State $ S_{i} \leftarrow $ compressed data size after adding $ b_i $ to $ \basebits $ \label{algo:greedyselect:line:S} \Comment{\autoref{eq:S}}
				
				\State $ \Delta_i' \leftarrow \Delta_i $ after adding $ b_i $ to $ \basebits $ \label{algo:greedyselect:line:deltaprime} \Comment{\autoref{eq:maxdevupdate}}
				
				\State $ \cost_{i} \leftarrow $ cost after adding $ b_i $ to $ \basebits $ \label{algo:greedyselect:line:cost} \Comment{\autoref{eq:cost}}
				
				\If{$ \cost_{i} < \cost_{\mathrm{loc}} $} \Comment{lowest cost bit in current iteration} \label{algo:greedyselect:line:mi_start}
					\State $ b_{\mathrm{loc}} \leftarrow b_i $
					\State $ \cost_{\mathrm{loc}} \leftarrow \cost_{i} $
				\EndIf \label{algo:greedyselect:line:mi_end}
				
			\EndFor \label{algo:greedyselect:line:iterate_m_end}
			
			\If{$ \cost_{\mathrm{loc}} > (1 + \alpha) \cdot \cost_{\mathrm{best}} $} \label{algo:greedyselect:line:termination} \Comment{early termination}
				\State \Return $ \basebits_{\mathrm{best}} $
			\EndIf
			
			\State Append $ b_{\mathrm{loc}} $ to $ \basebits $  \label{algo:greedyselect:line:updates_start}
			\State Expand \algoNameBaseTree{} with $ b_{\mathrm{loc}} $
			
			\If{$ \cost_{\mathrm{loc}} < \cost_{\mathrm{best}} $}  \Comment{update global best}
				\State $ \basebits_{\mathrm{best}} \leftarrow \basebits $
				\State $ \cost_{\mathrm{best}} \leftarrow \cost_{i} $
			\EndIf  \label{algo:greedyselect:line:updates_end}

		\EndWhile
		\State \Return $ \basebits_{\mathrm{best}} $
	\end{algorithmic}
	\\[-1em] \bottomrule
\end{tabular}

\end{figure}

As noted previously, selecting appropriate base bits is critical to the performance of GD.
For this task, we propose the \algoNameGreedySelect{} algorithm, which is designed to optimise compressed data analytics performance while maintaining good compression.
As outlined in Algorithm~\ref{algo:greedyselect}, \algoNameGreedySelect{} begins by adding all constant bits (i.e. bits that have the same value for all samples) to $ \basebits $ and expanding \algoNameBaseTree{} accordingly (lines~\ref{algo:greedyselect:line:constant_bits_start}--\ref{algo:greedyselect:line:constant_bits_end}).
It then iteratively adds bits to $ \basebits $ while minimising a cost function until termination.
Each iteration includes a local optimisation process followed by a global optimisation step.

In the local optimisation, \algoNameGreedySelect{} considers each column of the data and identifies the most significant bit, $ b_i $, that is currently mapped to the deviation (i.e. $ b_i \notin \basebits $, line~\ref{algo:greedyselect:line:nextmostsigbit}).
The number of bases, $ \nblocal $, resulting from adding $ b_i $ to $ \basebits $ is then computed using \algoNameBaseTree{} (line~\ref{algo:greedyselect:line:n_b}) as well as the resulting compressed data size, $ S_i $ (line~\ref{algo:greedyselect:line:S}).
The maximum deviation in column $ i $ if $ b_i $ is moved to the base, $ \Delta_i' $, is then computed (line~\ref{algo:greedyselect:line:deltaprime}) as:
\begin{align} \label{eq:maxdevupdate}
	\Delta_i' = \Delta_i \oplus 2^{b_i},
\end{align}
where $ \Delta_i $ is the current maximum deviation in column $ i $ and $ \oplus $ is the binary XOR operator.
Finally, the cost, $ \cost_i $, after adding $ b_i $ to $ \basebits $ is computed (line~\ref{algo:greedyselect:line:cost}) using the following cost function:
\begin{align} \label{eq:cost}
	\cost_i = \left(1 - \lambda \left( \Delta_{i}' / \maxdevzero \right) ^ 2 \right) \cdot S_i,
\end{align}
where $ \maxdevzero $ is the maximum deviation in dimension $ i $ after adding only the constant bits to $ \basebits $, and $ \lambda $ is known as the \textit{balancing factor} and defined such that $ 0 \le \lambda < 1 $.

The cost function in \autoref{eq:cost} is primarily dependent on $ S $, which favours high compression.
However, it is also scaled based on the maximum deviation such that it balances the number of bits added to $ \basebits $ from each dimension to improve analytics performance.
In the first iteration, scaling has no effect since $ \Delta_{i}' / \maxdevzero $ is $ 1 / 2 $ for all dimensions. % and $ \cost_i = (1 - \lambda / 4) \cdot S_i $
In subsequent iterations, the ratio $ \Delta_{i}' / \maxdevzero $ will be lower for dimensions which have had bits added to $ \basebits $.
Thus, the term $ (1 - \lambda ( \Delta_{i}' / \maxdevzero ) ^ 2 ) $ will be larger (resulting in higher cost) for dimensions that have had more bits added to $ \basebits $ and smaller (lower cost) for dimensions that have had fewer bits added to $ \basebits $.
The magnitude of this effect is controlled by the balancing factor,~$ \lambda $.
We found setting $ \lambda = 0.02 $ provided good results for most datasets.
At this setting, once 3--4 bits are added to $ \basebits $ from each dimension, the balancing effect becomes negligible and $ \cost_i \approx S_i $.
The local optimisation process is completed by keeping track of the lowest cost,~$ \cost_{\mathrm{loc}} $, and the corresponding bit, $ b_{\mathrm{loc}} $ (lines~\ref{algo:greedyselect:line:mi_start}--\ref{algo:greedyselect:line:mi_end}).

The first step in the global optimisation stage of \algoNameGreedySelect{} is to check the termination condition (line~\ref{algo:greedyselect:line:termination}).
Termination is controlled by the \textit{exploration factor},~$ \alpha > 0 $, which limits the amount of exploration beyond local minima.
The algorithm terminates if $ \cost_{\mathrm{loc}} $ exceeds $ 1 + \alpha $ times the current global best cost,~$ \cost_{\mathrm{best}} $.
We found setting $ \alpha = 0.1 $ performed well for most datasets.
If not terminated, \algoNameGreedySelect{} adds $ b_{\mathrm{loc}} $ to $ \basebits $, updates \algoNameBaseTree{} and, if required, updates $ \cost_{\mathrm{best}} $ (lines~\ref{algo:greedyselect:line:updates_start}--\ref{algo:greedyselect:line:updates_end}).
Overall, up to $ l_c d $ base bits combinations are examined ($ d $~per iteration and up to $ l_c $~iterations).

%The process of iteratively adding bits to $ \basebits $ is illustrated in \autoref{fig:algo_illustrations}(b) as a tree diagram.
%Each tree level corresponds to an iteration.
%The four child nodes at each level show $ \nblocal $ for each of the dataset's four dimensions.
%The chosen (lowest cost) bit at each iteration is highlighted in bold.
%Note that this may not correspond to the smallest $ \nblocal $ (e.g. level four).
%The global optimum (lowest cost) is highlighted in \textcolor{colour2}{red}.

By selecting base bits from most to least significant, \algoNameGreedySelect{} guarantees \textit{order preservation}~\cite{Hurst_2021}.
This means that for any two data samples, $ x_1 $ and $ x_2 $, and their corresponding bases, $ \mathrm{base}(x_1) $ and $ \mathrm{base}(x_2) $,
\begin{align}
	x_1 < x_2 \Rightarrow \mathrm{base}(x_1) \le \mathrm{base}(x_2).
\end{align}
Order preservation is critical for analytics performance.
Without it, results can be orders of magnitude worse.
Overall, by providing order preservation, balancing the number of base bits from each dimension and basing the cost function on compressed data size, \algoNameGreedySelect{} effectively balances the objectives of compressed data analytics and compression.

\subsection{Data preprocessing}
\label{sec:greedy_gd:preprocessing}

% Preprocessing illustration
\begin{figure}[!t]
	\centering
	\newcommand{\hlconst}[1]{\textcolor{colour1}{\textbf{#1}}}

\tikzset{
	preprocessing example style/.style={
		matrix of nodes,
		row sep=-\pgflinewidth,
		column sep=4\pgflinewidth,
		nodes={anchor=base east},
		anchor=north,
		minimum height=1.35em,
		font=\scriptsize,
	}
}

\footnotesize
\ttfamily
\begin{tikzpicture}
	% Without scaling
	\matrix [
		preprocessing example style,
	] (before)
	{
		 0.39 & \hlconst{0}0111110110001111010111000010100 \\
		37.83 & \hlconst{0}1000010000101110101000111101100 \\
		98.92 & \hlconst{0}1000010110001011101011100001010 \\
	};

	% With scaling
	\matrix [
		preprocessing example style,
		yshift=-1.4em,
	] (scaling) at (before.south)
	{
		   39. & \hlconst{0}1\hlconst{000}01000011100000000\hlconst{0000000000} \\
		3,783. & \hlconst{0}1\hlconst{000}10101101100011100\hlconst{0000000000} \\
		9,892. & \hlconst{0}1\hlconst{000}11000011010100100\hlconst{0000000000} \\
	};

	% Converted to integer
	\matrix [
		preprocessing example style,
		yshift=-1.4em,
	] (integer) at (scaling.south)
	{
		   39 & \hlconst{000000000000000000}00000000100101 \\
		3,783 & \hlconst{000000000000000000}00111011000111 \\
		9,892 & \hlconst{000000000000000000}10011010100100 \\
	};

	% Annotations
	\node[yshift=0.4em, font=\rmfamily, anchor=north] (labelBefore) at (before.south) {(a) Original data (1 constant bit)};
	\node[yshift=0.4em, font=\rmfamily, anchor=north] (labelScaling) at (scaling.south) {(b) After scaling (14 constant bits)};
	\node[yshift=0.4em, font=\rmfamily, anchor=north] (labelInteger) at (integer.south) {(c) Integer (18 constant bits)};
\end{tikzpicture}
\normalfont
	\vspace{0.0em}
	\caption{
		Selected data points from the first dimension of the \textit{Aarhus Citylab} dataset~\cite{AarhusKommune_2017} in decimal and IEEE-754~\cite{IEEE754_2019} floating point representation (a)~before and (b)~after scaling and (c)~as integers.
		Bits that are constant across the entire dataset are highlighted in \textcolor{colour1}{blue}.
	}
	\label{fig:illustration_preprocessing}
\end{figure}

Transforming data into a more amenable form is a typical step in compression algorithms.
In the case of GD, floating point data can be compressed significantly better by applying scaling and then converting to integers.
For example, consider the data illustrated in \autoref{fig:illustration_preprocessing}, which is from the \textit{Aarhus CityLab} dataset~\cite{AarhusKommune_2017}.
The original data, \autoref{fig:illustration_preprocessing}(a), has just two decimal places but only one constant bit. % uses nearly all bits in its floating point representation.
After scaling to remove the decimal places, \autoref{fig:illustration_preprocessing}(b), many more bits are constant, meaning that they can be assigned to the base by GD, improving compression performance.
Finally, converting the data to integer results in even more constant bits.

\subsection{Configuration on a data subset}
\label{sec:greedy_gd:sampling}

For large and complex datasets, configuring GD can still take substantial time.
%, even with of \basetreealgorithm{}.
In cases where this is prohibitive, we follow the suggestion of~\cite{Vestergaard_2020} to use a subset of the data.
However, we note that this introduces two important risks: 1)~the number of decimal places in the dataset could by greater than in the data subset, resulting in inaccurate preprocessing, and 2)~constant bits in the data subset may vary elsewhere in the dataset, which could result in violating order preservation.
In light of these risks, we propose using the full dataset to select preprocessing and determine constant bits and then run the remainder of \algoNameGreedySelect{} on a subset of the data.
%We examine the impact of this on compression performance in \autoref{sec:results:sampling}.

\subsection{Complexity}
\label{sec:greedy_gd:complexity}

The time complexity for configuring \newgdname{} (i.e. running \algoNameGreedySelect{}) is as follows.
First, the constant bits must be identified, which requires examining all $ n l_c $ bits.
Next, the base tree must be extended with these bits, but since they are constant, there is no need to split the tree and only nodes for the constant bits must be added.
In the worst case this is $ l_c $ (all bits constant).
The main while loop in \algoNameGreedySelect{} runs up to $ l_c $ times before all bits are in $ \basebits $.
For each iteration, the for loop runs $ d $ times.
Within the for loop, all steps are $ O(1) $ except for finding $ n_{b,i} $, which requires checking whether to split each node in \algoNameBaseTree{}.
In the worse case, this is $ O(n) $ since one bit needs to be checked across potentially every sample.
The global update step is constant time, except for updating \algoNameBaseTree{}, which is $ O(n) $ in the worse case.
Overall then, the while loop has a worst case time complexity of $ O(ndl_c) $.
Therefore, the total worst case time complexity for \newgdname{} configuration is $ O(nl_c + l_c + ndl_c) $ or $ O(ndl_c) $.
Noting that $ l_c $ is a multiple of $ d $ ($ 32d $ for single precision data and $ 64d $ for double precision data), we can simplify this to $ O(nd^2) $.
In practice, the while loop runs substantially fewer than $ l_c $ times, with the number of iterations reduced by both the number of constant bits and early termination.
The local optimisation loop may also have fewer iterations if all bits from one or more dimensions are added to the base.

Configuring \newgdname{} on a data subset can reduce its time complexity.
Assuming the full dataset is used for preprocessing and identification of constant bits and a data subset is used thereafter, then the time complexity of \algoNameGreedySelect{} is $ O(nl_c + l_c + ndrl_c) $ or $ O(nl_c(1 + dr))$, where $ 0 < r < 1 $ is the fraction of the data contained in the subset.
If $ {r < 1/d} $, then the worst case time complexity reduces to $ O(nd) $.

Compression is comparatively straightforward as it only involves splitting the base bits from the deviation bits and adding base IDs and counts.
Nonetheless, it requires accessing the full binary data and thus has time complexity $ O(nd) $.

	\section{Performance Evaluation}
\label{sec:results}

% Datasets table
\begin{table}[!t]
	\centering
	\renewcommand{\arraystretch}{1.2}
	\captionsetup{width=\linewidth}
	\caption{
		Datasets used for evaluating \newgdname{}.
	}
	\scriptsize
	\setlength\tabcolsep{5pt}
	\begin{tabular}{lllrrr}
	\toprule
	\textbf{Dataset}                                 & \textbf{Type} & \textbf{Precision} & $ \boldsymbol{n} $ & $ \boldsymbol{d} $ & \textbf{Size} (kB) \\ \midrule
	Aarhus Citylab~\cite{AarhusKommune_2017}         & float         & single             &             26,387 &                  4 &                422 \\
	Aarhus pollution 172156~\cite{Ali_2015}          & int           & single             &             17,568 &                  5 &                351 \\
	Aarhus pollution 204273~\cite{Ali_2015}          & int           & single             &             17,568 &                  5 &                351 \\
	Chicago beach water I~\cite{CityChicago_2022b}   & float         & single             &             39,829 &                  5 &                797 \\
	Chicago beach water II~\cite{CityChicago_2022b}  & float         & single             &             10,034 &                  6 &                241 \\
	Chicago beach weather~\cite{CityChicago_2022a}   & float         & single             &             86,694 &                  9 &              3,121 \\
	Chicago beach weather~\cite{CityChicago_2022a}   & int           & single             &             86,763 &                  5 &              1,735 \\
	Chicago taxi trips~\cite{CityChicago_2022c}      & float         & double             &          3,466,498 &                 10 &            277,320 \\
	CMU IMU acceleration~\cite{Torre_2009}           & float         & single             &            134,435 &                  3 &              1,613 \\
	CMU IMU velocity~\cite{Torre_2009}               & float         & single             &            134,435 &                  3 &              1,613 \\
	CMU IMU magnetic~\cite{Torre_2009}               & float         & single             &            134,435 &                  3 &              1,613 \\
	CMU IMU position~\cite{Torre_2009}               & float         & single             &            134,435 &                  4 &              2,151 \\
	CMU IMU all~\cite{Torre_2009}                    & float         & single             &            134,435 &                 13 &              6,991 \\
	COMBED mains power~\cite{Batra_2014}             & float         & double             &             82,888 &                  3 &                995 \\
	COMBED UPS power~\cite{Batra_2014}               & float         & double             &             86,199 &                  3 &              1,035 \\
	Melbourne city climate~\cite{CityMelbourne_2020} & float         & single             &             56,570 &                  3 &                679 \\
	Gas turbine emissions~\cite{Kaya_2019}           & float         & single             &             36,733 &                 11 &              1,616 \\
	Household power usage~\cite{Hebrail_2012}        & float         & single             &          2,049,280 &                  7 &             57,380 \\ \bottomrule
\end{tabular}

	\label{tab:datasets}
\end{table}

We evaluated the performance of \newgdname{} in terms of compression (\autoref{sec:results:cr}), compressed data analytics (\autoref{sec:results:analytics}), configuration runtime (\autoref{sec:results:runtime}) and configuration using data subsets (\autoref{sec:results:sampling}).
In total, 18~datasets were used for evaluation (\autoref{tab:datasets}), which include a range of sizes, dimensionalities, data types and precisions.
Note that non-numerical attributes and rows with missing values were omitted.
The \textit{Chicago beach weather} dataset was also split between float and integer columns and some derived fields were omitted from the \textit{Chicago taxi trips} dataset.
All evaluations were performed on a laptop with an Intel Core i7-10510U 1.80 GHz CPU and 16 GB of RAM using Python~3.8.10.

%\todo[inline]{Re-write the below, and part of related work, if I add other compressors with direct compressed data analytics capabilities Related work: Lossless compressors for numerical data that provide direct compressed data analytics capabilities are few and far between. To the best of our knowledge there are only... More options are available in the lossy compression space, for example....}

First and foremost, \newgdname{} was compared to previous versions of GD, namely \gdINFOCOM{}~\cite{Vestergaard_2020} and \gdGLEAN{}~\cite{Hurst_2022}.
Note that, to apply \gdINFOCOM{} to multidimensional data, it was necessary to modify its termination criteria such that it explores slightly beyond local minima (similar to \newgdname{}).
We also compared to enhanced version of \gdINFOCOM{} and \gdGLEAN{} that used preprocessing and counted bases using \algoNameBaseTree{}.
We refer to these compressors as \gdINFOCOMplus{} and \gdGLEANplus{}, respectively.
To use \algoNameBaseTree{}, it was necessary to reverse the order of iteration so that \gdINFOCOMplus{} starts with $ \basebits = \emptyset $ and iteratively adds bits, rather than starting with all bits in $ \basebits $ and iteratively removing bits.
Throughout our evaluation, we set $ \alpha = 0.1 $ and $ \lambda = 0.02 $ for \newgdname{}.

% Main CR figure
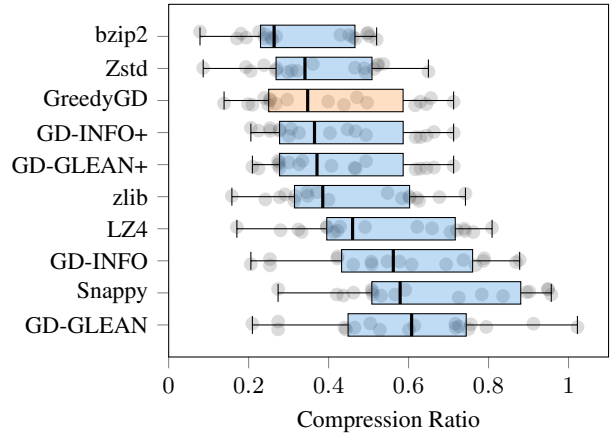
\begin{figure}[!t]
	\centering
	\begin{tikzpicture}
    \begin{axis}[
        % General options
        single boxplot style,
        width=0.45\textwidth, height=0.38\textwidth,
        xmax=1.1, xmin=0,
        ymax=11, ymin=0,
        % Ensure scatter plot renders below box plot
        set layers=standard,
        mark layer=axis ticks,
        % Labels
        ytick={1,...,10},
        xtick={0, 0.2, 0.4, 0.6, 0.8, 1.0},
        yticklabels={
            bzip2, Zstd, \newgdname, \gdINFOCOMplus, \gdGLEANplus, zlib, LZ4, \gdINFOCOM, Snappy, \gdGLEAN
        },
        xlabel={Compression Ratio},
        y dir=reverse,  % reverse order of y axis
        ]
        % Scatter plots
        \foreach \i in {2,...,11} {
            \addplot+ [
            only marks,
            mark=*,
            black!70,
            mark options={draw opacity=0.0, fill opacity=0.2, scale=1.3},
            ] table[x index=\i, y index=\the\numexpr\i+10] {\dataResultsCompression};
        }

        % Box plots
        %\pgfplotsset{cycle list shift=-1}
        \foreach \i in {2,...,11} {
            \ifthenelse{\equal{\i}{4}}
            {% True case
                \addplot+ [boxplot, fill=colour2] table[y index=\i] {\dataResultsCompression};
            }
            {% false case
                \addplot+ [boxplot] table[y index=\i] {\dataResultsCompression};
            }
        }
    \end{axis}
\end{tikzpicture}
	\vspace{-0.4em}
	\caption{
		Box plots of CR for each compressor ordered from best (top) to worst (bottom) based on median CR.
	}
	\label{fig:cr}
	\vspace{-0.4em}
\end{figure}

\subsection{Compression Ratio}
\label{sec:results:cr}

% q-q plots of compression performance
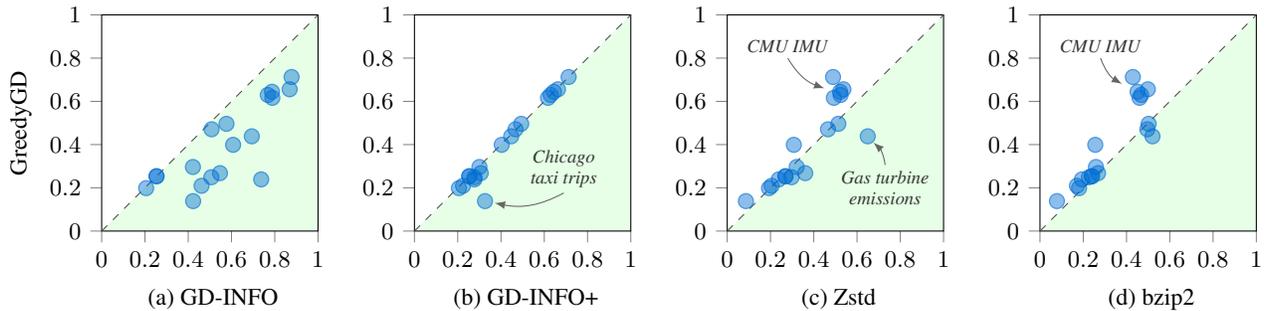
\begin{figure*}[!t]
	\centering
	\hspace{2.8em}  % This compensates for trimming the plot when computing the bounding box to get the caption center aligned
	\subfloat[\gdINFOCOM{}]{\begin{tikzpicture}[trim axis left, trim axis right]
	\begin{axis}[
		custom scatter plot style,
		width=0.27\textwidth, height=0.27\textwidth,
		xmin=0, xmax=1.0,
		ymin=0, ymax=1.0,
		ylabel={\newgdname{}},
		set layers=standard,
	]
		% Fill plot
		\begin{pgfonlayer}{axis background}    % select the background layer
			\fill[fill=green!10] (0,0)--(1,0)--(1,1);
		\end{pgfonlayer}
		% Plots
		\addplot+ [] table[x index=9, y index=4] {\dataResultsCompression};
		\addplot [sharp plot, mark=none, dashed, black!70] coordinates {(0, 0) (1, 1)};
	\end{axis}
\end{tikzpicture}}
	\hfill
	\subfloat[\gdINFOCOMplus{}]{\begin{tikzpicture}[trim axis left, trim axis right]
	\begin{axis}[
		custom scatter plot style,
		width=0.27\textwidth, height=0.27\textwidth,
		xmin=0, xmax=1.0,
		ymin=0, ymax=1.0,
		set layers=standard,
	]
		% Fill plot
		\begin{pgfonlayer}{axis background}    % select the background layer
			\fill[fill=green!10] (0,0)--(1,0)--(1,1);
		\end{pgfonlayer}
		% Add plots
		\addplot+ [] table[x index=5, y index=4] {\dataResultsCompression};
		\addplot [sharp plot, mark=none, dashed, black!70] coordinates {(0, 0) (1, 1)};
		% Annotate dataset
		\node[black!80, anchor=west, font=\scriptsize, inner sep=0.2em, align=center] (dataset) at (0.51, 0.29) {\textit{Chicago}\\\textit{taxi trips}};
		\draw [black!60, -latex, solid] (dataset.south) to [bend left=20] (0.37, 0.13);
	\end{axis}
\end{tikzpicture}}
	\hfill
	\subfloat[Zstd]{\begin{tikzpicture}[trim axis left, trim axis right]
	\begin{axis}[
		custom scatter plot style,
		width=0.27\textwidth, height=0.27\textwidth,
		xmin=0, xmax=1.0,
		ymin=0, ymax=1.0,
		set layers=standard,
	]
		% Fill plot
		\begin{pgfonlayer}{axis background}    % select the background layer
			\fill[fill=green!10] (0,0)--(1,0)--(1,1);
		\end{pgfonlayer}
		% Add plots
		\addplot+ [] table[x index=3, y index=4] {\dataResultsCompression};
		\addplot [sharp plot, mark=none, dashed, black!70] coordinates {(0, 0) (1, 1)};
		% Lower outlier
		\node[black!80, anchor=west, font=\scriptsize, inner sep=0.2em, align=center] (dataset) at (0.5, 0.2) {\textit{Gas turbine}\\\textit{emissions}};
		\draw [black!60, -latex, solid] (dataset.north) to [bend right=20] (0.675, 0.41);
		% Upper outliers
		\node[black!80, anchor=south, font=\scriptsize, inner sep=0.2em] (dataset2) at (0.28, 0.8) {\textit{CMU IMU}};
		\draw [black!60, -latex, solid] (dataset2.south) to [bend right=20] (0.45, 0.66);
	\end{axis}
\end{tikzpicture}}
	\hfill
	\subfloat[bzip2]{\begin{tikzpicture}[trim axis left, trim axis right]
	\begin{axis}[
		custom scatter plot style,
		width=0.27\textwidth, height=0.27\textwidth,
		xmin=0, xmax=1.0,
		ymin=0, ymax=1.0,
		set layers=standard,
	]
		% Fill plot
		\begin{pgfonlayer}{axis background}    % select the background layer
			\fill[fill=green!10] (0,0)--(1,0)--(1,1);
		\end{pgfonlayer}
		% Add plots
		\addplot+ [] table[x index=2, y index=4] {\dataResultsCompression};
		\addplot [sharp plot, mark=none, dashed, black!70] coordinates {(0, 0) (1, 1)};
		% Annotations
		\node[black!80, anchor=south, font=\scriptsize, inner sep=0.2em] (dataset2) at (0.28, 0.8) {\textit{CMU IMU}};
		\draw [black!60, -latex, solid] (dataset2.south) to [bend right=20] (0.38, 0.68);
	\end{axis}
\end{tikzpicture}}
	\vspace{0.4em}
	\caption{
		Pairwise comparison of compression ratios for \newgdname{} and selected compressors across all 18 datasets with noteworthy datasets annotated.
		Points below (above) the diagonal indicate better (worse) compression with \newgdname{}.
	}
	\label{fig:cr_pairwise_comparisons}
\end{figure*}

The compression performance of \newgdname{} was compared to the other GD versions mentioned above, as well as five widely-used universal compressors, namely zlib~\cite{Deutsch_1996a}, Zstd~\cite{Facebook_}, Snappy~\cite{Google_2021}, bzip2~\cite{Seward_2019} and LZ4~\cite{Collet_2020}.
% Other citations for universal compressors: DEFLATE~\cite{Deutsch_1996}, Zstd~\cite{Szorc_2020}, Snappy~\cite{Moreira_2021}
All universal compressors were run in one-shot mode at maximum compression.
%, meaning that they may have used higher runtime and/or memory.
%(except for Snappy, which does not offer compression level tuning).
The distributions of compression ratios (CRs) resulting from applying these compressors to all 18 datasets are illustrated in \autoref{fig:cr}, where grey dots indicate CR values for individual datasets.
As can be seen, \newgdname{} is competitive with the best of the universal compressors and outperforms all other GD versions.
%Median CR values are also listed in the first column of \autoref{tab:results:analytics}.

%As can be seen, \newgdname{} (median CR 0.348) outperforms zlib (0.385), LZ4 (0.460), Snappy (0.579) and all other GD compressors (\gdINFOCOM{}: 0.562, \gdGLEAN{}: 0.608) and nearly matches the performance of Zstd (0.341).

\autoref{fig:cr_pairwise_comparisons} provides more granulated comparisons.
Each point in these figures represents the CR for a single dataset with \newgdname{} (y-axis) and another compressor (x-axis).
Points below the diagonal therefore represent better performance with \newgdname{}.
As can be seen in \autoref{fig:cr_pairwise_comparisons}(a), \newgdname{} substantially outperforms \gdINFOCOM{}, which is the existing state-of-the-art version of GD in terms of compression.
Indeed, with \newgdname{} having a median CR of 0.348 compared to \gdINFOCOM{}'s 0.562, \newgdname{} provides 1.6$ \times $ more compression on average.
\autoref{fig:cr_pairwise_comparisons}(b), which compares \newgdname{} to \gdINFOCOMplus{}, shows that, even with data preprocessing, \gdINFOCOM{} never outperforms \newgdname{}.
\autoref{fig:cr_pairwise_comparisons}(c) and (d) compare \newgdname{} to Zstd and bzip2, respectively.
As can be seen, \newgdname{} has very similar CR for most datasets compared to these compressors, with the exception of the IMU data.
In particular, \newgdname{} has very similar performance to Zstd, which has a median CR of 0.341.
Note that GD can achieve even more compression if applying entropy coding to the base IDs, however, this compromises random access and thereby analytics efficiency~\cite{Vestergaard_2020}.

\subsection{Analytics on compressed data}
\label{sec:results:analytics}

% q-q plots of AR
% Joint AR vs. ADR distribution
\begin{figure*}[!t]
    \centering
    \begin{minipage}[t]{.6\textwidth}
        \vspace{-1em}
        \hfill
        \subfloat[\gdGLEAN{}]{\newcommand{\minvalue}{0.6}
\newcommand{\maxvalue}{2.5}

\begin{tikzpicture}[trim axis left, trim axis right]
	\begin{axis}[
		custom scatter plot style,
		width=0.48\textwidth, height=0.48\textwidth,
		xmin=\minvalue, xmax=\maxvalue,
		ymin=\minvalue, ymax=\maxvalue,
		ylabel={\newgdname{}},
		set layers=standard,
	]
		% Fill plot
		\begin{pgfonlayer}{axis background}    % select the background layer
			\fill[fill=green!10] (\minvalue,\minvalue)--(\maxvalue,\minvalue)--(\maxvalue,\maxvalue);
		\end{pgfonlayer}

		% Add plots
		\addplot+ [] table[x index=0, y index=4] {\dataResultsClusteringAR};
		\addplot [sharp plot, mark=none, dashed, black!70] coordinates {(0, 0) (10, 10)};

		% Annotate dataset
		\node[black!80, anchor=south west, font=\scriptsize, inner sep=0.2em, align=center] (dataset) at (0.7, 2.05) {\textit{Gas turbine emissions}\\\textit{\& COMBED UPS}};
		\draw [black!60, -latex, solid] (dataset.south) to (1.95, 1.55);
	\end{axis}
\end{tikzpicture}}
        \hfill
        \subfloat[\gdGLEANplus{}]{\newcommand{\minvalue}{0.6}
\newcommand{\maxvalue}{2.5}

\begin{tikzpicture}[trim axis left, trim axis right]
	\begin{axis}[
		custom scatter plot style,
		width=0.48\textwidth, height=0.48\textwidth,
		xmin=\minvalue, xmax=\maxvalue,
		ymin=\minvalue, ymax=\maxvalue,
		set layers=standard,
	]
		% Fill plot
		\begin{pgfonlayer}{axis background}    % select the background layer
			\fill[fill=green!10] (\minvalue,\minvalue)--(\maxvalue,\minvalue)--(\maxvalue,\maxvalue);
		\end{pgfonlayer}

		% Add plots
		\addplot+ [] table[x index=1, y index=4] {\dataResultsClusteringAR};
		\addplot [sharp plot, mark=none, dashed, black!70] coordinates {(0, 0) (10, 10)};

		% Annotate dataset
		\node[black!80, anchor=west, font=\scriptsize, inner sep=0.2em, align=center] (dataset) at (0.7, 2) {\textit{Aarhus Citylab}};
		\draw [black!60, -latex, solid] (dataset.east) to  [bend left=20] (2.0, 1.2);
	\end{axis}
\end{tikzpicture}}
        \hfill
        \vspace{0.4em}
        \caption{
            Pairwise comparison of AR for $ k $-means clustering on compressed data using \newgdname{} and (a) \gdGLEAN{} and (b) \gdGLEANplus{} across all 18 datasets.
            Points below (above) the diagonal indicate better (worse) analytics performance with \newgdname{}.
        }
        \label{fig:ar_pairwise_comparisons}
    \end{minipage}
    \hfill
    \begin{minipage}[t]{.38\textwidth}
        \vspace{0pt}
        \centering
        \begin{tikzpicture}
	\begin{semilogxaxis}[
		custom scatter plot style,
		width=0.99\textwidth, height=0.78\textwidth,
		xmin=0.001, xmax=0.6,
		ymin=0.75, ymax=2.4,
		ylabel={Approximation Ratio},
		xlabel={Analytics Data Ratio},
		legend columns=1,
		legend image post style={scale=0.8},
		legend style={
			font=\scriptsize,
			/tikz/every even column/.append style={column sep=0.3em},
			/tikz/every odd column/.append style={column sep=0.1em},
			anchor=north east,
			legend pos=north east,
		},
		set layers=standard,
	]
		% Add scatter plots
		\addplot+ [] table[x index=4, y index=2] {\dataResultsClusteringADRGLEAN};
		\addplot+ [] table[x index=4, y index=2] {\dataResultsClusteringADRGLEANplus};
		\addplot+ [] table[x index=4, y index=2] {\dataResultsClusteringADRGreedy};

		% Highlight worst datasets from GLEAN
		\pgfplotsset{cycle list shift=-3}
		\addplot+ [mark options={draw=black, fill opacity=0, scale=1.4, line width=0.8pt}]coordinates {(0.012248, 2.149351) (0.023665, 2.139292) (0.089023, 1.580823)};
		\addplot+ [mark options={draw=black, fill opacity=0, scale=1.3, line width=0.8pt}]coordinates {(0.003833, 1.004989) (0.041452, 1.372720) (0.008500, 1.079832)};
		\addplot+ [mark options={draw=black, fill opacity=0, scale=1.8, line width=0.8pt}]coordinates {(0.006479, 1.000049) (0.017771, 1.348067) (0.004517, 1.010283)};

		% Annotate AR=1
		\begin{pgfonlayer}{axis background}    % select the background layer
			\fill[fill=black!10] (0.001,0.75)--(0.001,1)--(0.6,1)--(0.6,0.75);
		\end{pgfonlayer}

		% Legend
		\legend{\gdGLEAN, \gdGLEANplus, \newgdname}
	\end{semilogxaxis}
\end{tikzpicture}
        \caption{
            Joint distribution of analytics data ratio (ADR) and clustering approximation ratio (AR) for \newgdname{} and both \gdGLEAN{} versions across all datasets.
            The three datasets with the worst performance under \gdGLEAN{} are highlighted with a black border for all three compressors.
        }
        \label{fig:adr_vs_ar}
    \end{minipage}
\end{figure*}

% Analytics results table
\begin{table}[!t]
	\centering
	\renewcommand{\arraystretch}{1.1}
	\caption{
		Summary of compression \& analytics results.
		Best (second best) performance shown in bold (italics).
	}
	%	\scriptsize
	\begin{tabular}{lccccc}
	\toprule
	\textbf{Compressor} &  \textbf{CR}   &  \textbf{ADR}  &  \textbf{AR}   &  \textbf{AMI}  & \textbf{Silhouette} \\ \midrule
	\newgdname{}        & \textbf{0.348} & \textit{0.011} &     1.109      & \textbf{0.758} &   \textit{0.309}    \\
	\gdINFOCOMplus{}    & \textit{0.365} & \textbf{0.009} &     1.246      &     0.630      &        0.294        \\
	\gdGLEANplus{}      &     0.371      &     0.013      & \textbf{1.097} & \textit{0.753} &        0.295        \\
	\gdINFOCOM{}        &     0.562      &     0.014      &     1.806      &     0.368      &        0.162        \\
	\gdGLEAN{}          &     0.608      &     0.048      & \textit{1.099} &     0.639      &   \textbf{0.313}    \\ \bottomrule
\end{tabular}

	\label{tab:results:analytics}
\end{table}

Approximations to various analytics tasks can be obtained directly from GD-compressed data without decompressing it.
This is done by performing analytics on the bases weighted by their counts~\cite{Hurst_2021}.
A significant benefit of this approach is that only a fraction as many points need to be analysed, since the number of bases is typically far less than the number of samples, i.e. $ n_b << n $.
This translates into savings in runtime and memory~\cite{Hurst_2022}.
Thus, to evaluate the effectiveness of compressed data analytics, we consider both the \textit{quality} of approximate analytics results and the \textit{amount} of data that needs to be accessed to perform analytics.

The median amount of data needed for analytics for each GD-based compressor, expressed via the analytics data ratio (ADR), is shown in \autoref{tab:results:analytics}.
As can be seen, \newgdname{} requires a mere 1.1\% of the uncompressed data to perform analytics, which is less than a quarter of that required by \gdGLEAN{} and 15\% less than \gdGLEANplus{}.

Existing GD compressors have been shown to support high clustering performance~\cite{Hurst_2021,Hurst_2022}.
We have thus thus used $ k $-means clustering~\cite{Hartigan_1979} to evaluate \newgdname{}'s analytics performance.
Our methodology mimicked that of~\cite{Hurst_2022}, such that we computed cluster centres directly from the bases and used these to cluster the original data points.
Clusterings of the original data were then used to compute Silhouette coefficients and then compared to clustering performed on uncompressed data to compute AR and AMI.
Note that each clustering was repeated 10 times with 100 initialisations each to minimise variance and sampling with $ n = \text{10,000} $ was used to compute the Silhouette coefficient to avoid excessive runtime.
The median results across all 18 datasets are presented in \autoref{tab:results:analytics}.
As can be seen, \newgdname{} provides best or near-best performance in each clustering performance metric.

The performance of \newgdname{} in comparison to \gdGLEAN{} and \gdGLEANplus{} was also examined more closely.
\autoref{fig:ar_pairwise_comparisons}(a) and (b) show the AR for each dataset using \newgdname{} (y-axis) compared to \gdGLEAN{} and \gdGLEANplus{}, respectively (x-axis).
As can be seen, \newgdname{} generally delivers similar or better performance and has a more reliable distribution of AR values with no outliers.

Overall, \newgdname{} provides solid performance across all analytics-related metrics we tested.
To illustrate this more holistically, we also present \autoref{fig:adr_vs_ar}, which plots AR against ADR for \newgdname{}, \gdGLEAN{} and \gdGLEANplus{} across all datasets.
In this graph, lower and further to the left corresponds to better performance.
As can be seen, the distribution of AR/ADR values for \newgdname{} is the most optimal.
The three datasets with the worts performance under \gdGLEAN{} are also highlighted with black borders for each compressor.
Through this, it can also be seen that both AR and ADR improve from \gdGLEAN{} to \gdGLEANplus{} and from \gdGLEANplus{} to \newgdname{}.

%%%%%%%%%%%%%%%%%%%%%%%%%%%%%%%%%%%%%%%%%%%%%%%%%%%%%%%%%%%%%%%%%%%%%%%%%%%%%

\subsection{Configuration runtime}
\label{sec:results:runtime}

% Runtime & sampling figures
\begin{figure*}[!t]
    \centering
    \begin{minipage}[t]{.315\textwidth}
        \vspace{0pt}
        \centering
        \begin{tikzpicture}
	\begin{axis}[
		width=\textwidth, height=0.9\textwidth,
		xmin=-0.7, xmax=3.7,
		ymin=-0.3, ymax=7,
		% Plot design
		ybar,
		bar width=0.6,
		xtick pos=bottom,
		% Labels
		ylabel={Runtime (seconds)},
		xtick={0, 1, 2, 3},
		xticklabels={\gdINFOCOM, \gdINFOCOMplus, \gdGLEANplus, \newgdname},
		x tick label style={rotate=30,anchor=north east},
%		minor y tick num=1,
	]
		\addplot [
			fill=colour1!50,
			fill opacity=0.6,
			draw=colour1!70,
			line width=0.5pt,
			error bars/.cd,
			y dir=both,
			y explicit,
			error bar style={line width=0.8pt},
			error mark options={draw=colour1!70, line width=8pt, mark size=0.3pt},
		] coordinates {
			(0,5.341) += (0,1.184529) -= (0,0.181312)
			(1,0.452) += (0,0.135249) -= (0,0.018950)
			(2,0.463) += (0,0.089077) -= (0,0.019632)
			(3,0.475) += (0,0.035897) -= (0,0.021138)
		};
	
		% Annotations
		\node[colour3, anchor=north west, font=\scriptsize, inner sep=0.2em, align=left] (basetree) at (0.57, 5.341) {Significant speed-up\\from \texttt{BaseTree}};
		\draw [colour3, -latex, solid, line width=0.8pt] (0.5, 5.341) to (0.5, 0.452);
		
		\draw [black!70, densely dotted, line width=0.5pt] (-0.5, 0.452) to (3.5, 0.452);

%		\node[colour4, anchor=north west, font=\scriptsize, inner sep=0.2em, align=center] (increase) at (1.4, 2.65) {Slight increase\\from optimisation};
%		\draw [colour4, -latex, solid, line width=0.5pt] (2.5, 0.0) to (2.5, 0.452);
%		\draw [colour4, -latex, solid, line width=0.5pt] (2.5, 1.0) to (2.5, 0.475);
	\end{axis}
\end{tikzpicture}
        \vspace{-0.6em}
        \caption{
            Configuration runtime for GD-based compressors for the \textit{COMBED mains power} dataset.
        }
        \label{fig:gd_config_runtime_comparison}
    \end{minipage}
    \hfill
    \begin{minipage}[t]{.315\textwidth}
        \vspace{0pt}
        \centering
        \begin{tikzpicture}
	\begin{axis}[
		custom line plot style,
		width=\textwidth, height=0.9\textwidth,
		xmin=0, xmax=12,
		ymin=0, ymax=0.75,
		ylabel={Runtime (seconds)},
		xlabel={No. dimensions},
		xtick={0,2,4,6,8,10,12},
	]
		% Add plots
		\addplot+ [
			error bars/.cd,
			x dir=both, x explicit,
			y dir=both, y explicit
		] table[x index=0, y index=1, y error minus index=2, y error plus index=3] {\dataResultsRuntimeVsDimensionality};
	\end{axis}
\end{tikzpicture}
        \vspace{-0.4em}
        \caption{
            \newgdname{} configuration runtime versus dimensionality averaged over subsamples of the dimensions of the \textit{Gas turbine emissions} dataset.
        }
        \label{fig:configuration_runtime_dimensionality}
    \end{minipage}
    \hfill
    \begin{minipage}[t]{.315\textwidth}
        \vspace{0pt}
        \centering
        \begin{tikzpicture}
	\begin{semilogxaxis}[
		custom line plot style,
		width=\textwidth, height=0.9\textwidth,
		ymin=0, ymax=1.65,
		enlarge x limits=0.1,
		ylabel={Median CR},
		xlabel={Data subset size},
		% Legend
		legend columns=1,
		legend image post style={scale=0.8},
		legend style={
			row sep=0em,
			font=\scriptsize,
			/tikz/every even column/.append style={column sep=0.3em},
			/tikz/every odd column/.append style={column sep=0.1em},
			anchor=north east,
			legend pos=north east,
			fill opacity=0.7,
		},
	]
		% Add plots
		\foreach \i in {3,4,1,2,5} {
			\addplot+ [
				sharp plot,
			] table[x index=0, y index=\i] {\dataResultsSamplingCR};
		}

		\legend{\gdINFOCOM, \gdINFOCOMplus, \gdGLEAN, \gdGLEANplus, \newgdname}
	\end{semilogxaxis}
\end{tikzpicture}
        \vspace{0.5em}
        \caption{
            Median full dataset CR versus size of the training set used to configure each compressor.
        }
        \label{fig:gd_config_sample_compression}
    \end{minipage}
\end{figure*}

The proposed \algoNameBaseTree{} algorithm significantly improves configuration runtime for GD.
\autoref{fig:gd_config_runtime_comparison} shows the median configuration runtime for the \textit{COMBED mains power} dataset over 50 trials for selected compressors.
Error bars indicate the minimum and maximum runtimes across the 50 trials.
As can be seen, applying \algoNameBaseTree{} results in a significant speed-up.
Indeed, comparing \gdINFOCOM{} (5.341~s) with \gdINFOCOMplus{} (0.452~s) indicates that \algoNameBaseTree{} provides a speed-up of 11.8$ \times $.
\gdGLEANplus{} (0.463~s) and \newgdname{} (0.475~s) are slightly slower than \gdINFOCOMplus{} due to their optimisations for analytics, but still deliver speed-ups of 11.5$ \times $ and 11.2$ \times $, respectively.
Given the dataset size of 995~kB, \newgdname{}'s runtime corresponds to a throughput of 2.1~MB/s.
Note that our implementation is not optimised for speed and is written in Python, so the absolute runtime values should not be taken as indicative of the algorithms' capabilities.

As stated in \autoref{sec:greedy_gd:complexity}, \newgdname{}'s worst case configuration time complexity is $ O(nd^2) $.
However, as was noted, typical runtimes should be substantially faster.
To examine the relationship between configuration runtime and dimensionality, we measured the time to configure \newgdname{} on random subsets of the columns of the \textit{Gas turbine emissions} dataset.
For each dimensionality (1~to~11), we selected 50 random combinations of the dataset's columns (apart from $ d = $~1, 10 and 11 where there are fewer than 50 combinations) and ran 10 trials on each.
The median runtime across all combinations and trials for each dimensionality is shown in \autoref{fig:configuration_runtime_dimensionality}.
Error bars indicate the 5$ ^{\text{th}} $ and 95$ ^{\text{th}} $ percentiles.
As can be seen, \newgdname{} scales much better than quadratic in practice.
That is, the runtime for $ d = $~11 (0.638~s) is only 16.4 times that for $ d = $~1 (0.039~s), which is close to linear instead of quadratic.

\subsection{Configuration on data subsets}
\label{sec:results:sampling}

The effect of configuring GD on a data subset was evaluated in terms of CR for a range of subset sizes from 10 to 10,000 samples.
Subsets were chosen randomly and, as proposed in \autoref{sec:greedy_gd:sampling}, data preprocessing and constant bits were determined from the entire dataset.
The median CR across all datasets for each subset size is shown in \autoref{fig:gd_config_sample_compression}.
As can be seen, \newgdname{} and \gdINFOCOMplus{} exhibit very stable compression ratios across all subset sizes.
Indeed, even with a mere 250 samples, \newgdname{} delivers a median CR of 0.368, which is only 5.7\% worse than if the entire dataset is used for configuration.
If 10,000 samples are used, the median CR for \newgdname{} is only 1.4\% worse compared to configuration on the full dataset at 0.353.

	\section{Conclusion}
\label{sec:conclusion}

In this paper, we have proposed a new GD-based compression algorithm, \newgdname{}, that is designed for high accuracy direct compressed data analytics, while maintaining good compression.
The proposed scheme shows improvement in analytics performance compared to existing GD methods, while also running considerably faster and delivering even better compression.
We have also demonstrated that \newgdname{} can be reliably configured on very small data subsets without compromising significantly on compression.
In future work, we intend to investigate using \newgdname{} for online compression in an IoT network and additional analytics tasks.

	\section{Acknowledgements}

This work is supported by the Analytics Straight on Compressed IoT Data (Light-IoT) project (Grant No. 0136-00376B), granted by the Danish Council for Independent Research, and Aarhus University's DIGIT Centre.

	\bibliographystyle{IEEEtran}
	\bibliography{IEEEabrv,ms}
\end{document}